%% file: main.tex
\pdfoutput=1
\documentclass[lettersize,journal]{IEEEtran}
\usepackage{amsmath}
\usepackage{amssymb}
\usepackage{bm}
\usepackage{graphicx}
\usepackage{array}
\usepackage{cite}
\usepackage[bookmarks=false,hidelinks,hypertexnames=false]{hyperref}
\usepackage{xcolor}
\usepackage{amsthm}

\newtheorem{proposition}{Proposition}

\theoremstyle{remark}
\newtheorem{remark}{Remark}

\newcounter{finding}
\newcommand{\finding}[1]{\refstepcounter{finding}\label{#1}\emph{Finding~\thefinding}}

\newcommand{\vect}[1]{\bm{#1}}
\newcommand{\Ex}{\mathbb{E}}

\newcommand{\CN}{\mathcal{CN}}
\newcommand{\Ith}{I_{\mathrm{th}}}

\newcommand{\alphac}{\alpha_{c}}

\newcommand{\SNR}{\mathrm{SNR}}
\newcommand{\vp}{\vect{p}}

\newcommand{\veta}{\vect{\eta}}
\newcommand{\Rtotal}{R_{\mathrm{total}}}

\title{Positioning and Radiation Efficiency Degrees of Freedom in
       Cascaded Pinching-Antenna Arrays}

\author{Ozgur Ergul
\thanks{This work has been submitted to the IEEE for possible publication. Copyright may be transferred without notice, after which this version may no longer be accessible.}
\thanks{The author is with the Department of Electrical - Electronic Engineering, Faculty of Engineering, Gazi University, Ankara, Turkiye (e-mail: ergul@gazi.edu.tr).}}

\begin{document}

\maketitle
\IEEEpeerreviewmaketitle

\input{body_abstract}

\input{body_section_I_introduction}

\input{body_section_II_system_model}

\input{body_section_III_problem_formulation}

\input{body_section_V_simulation_results}

\input{body_section_VII_conclusion}

\input{hardcoded_bibliography}
\end{document}

%% file: body_abstract.tex
\begin{abstract}
In a pinching-antenna system (PASS), a single source feeds multiple pinches, resulting in a cascade that couples their radiated powers. A shared radiated-power limit, such as the interference power constraint in an underlay cognitive radio network, couples them again. We study such a network where PASS serves the secondary users. First, we formulate the joint optimization problem of the two degrees of freedom (DoF), pinch positions and radiation efficiencies, 
and characterize the single-link optimum analytically. Then, rather than proposing one method, we present the results for a grid of schemes whose two axes are the two DoF. 
Under an order-free attribution, positioning is worth $1.34\times$ radiation efficiency, and the two are near-perfect substitutes. We also show that substitutability is due to the constraint and not the cascade, since relaxing the constraint by $20$\,dB removes $97\%$ of their interaction. 
At the joint reference, not knowing the primary receiver (PR) location costs $6.23$\,bit/s/Hz, more than twice the $2.87$ that optimizing both DoF is worth.
Instantaneous fading information adds a further $0.29$\,bit/s/Hz.
Allocating efficiencies at equally spaced positions reaches $98\%$ of the joint reference at about three orders of magnitude less computation. 
\end{abstract}

\begin{IEEEkeywords}
Pinching antennas, flexible-antenna positioning, degrees of freedom, cascade power model, cognitive radio.
\end{IEEEkeywords}

%% file: body_section_I_introduction.tex
\section{Introduction}
\label{sec:introduction}

\IEEEPARstart{P}{inching} antennas, first demonstrated by NTT DOCOMO~\cite{NTT2024}, route signals through a dielectric waveguide and radiate them wherever a separate dielectric, a \emph{pinch}, is clipped onto it.
Pinching-antenna systems (PASS) were introduced and analyzed in~\cite{Ding2025Flex} and a survey is presented in~\cite{Yang2026PASS}. 
Each pinch is a low-gain antenna whose proximity to a user reduces free-space path loss. Choosing the position of these pinches provides one degree of freedom (DoF), and another can be obtained by choosing the fraction of the incoming power to radiate at each pinch, its radiation efficiency. This forms a cascade where the extracted power at an upstream pinch limits the power of the downstream pinches, coupling these pinches. Any shared limit on the radiated power couples the pinches again. The limit may be imposed to restrict interference on another nearby system. In this paper, we characterize what the two DoF are worth, separately and together, under a shared budget. One central finding is that the second coupling is what makes them interchangeable, not the cascade.

Previous studies on PASS mostly focus on optimizing pinching beamforming for sum or max-min rate under an idealized feed, neglect in-waveguide attenuation and assume an equal power split among the pinches~\cite{Zhao2026TCOM, Zhao2026TWC}. 
Analytical placement rules for a single pinch, derived in~\cite{Ding2025Placement}, rest on the same lossless model, which~\cite{Ding2025Blockage} also adopts in its analysis of line-of-sight (LOS) blockage; \cite{Xu2026Atten} extends the single-pinch rule to in-waveguide attenuation under free-space path loss.
Recent studies examine the cascade structure. In~\cite{Vashakidze2025NOMA}, sequential extraction with waveguide attenuation is introduced, where the per-pinch powers are optimized under a total-power constraint. Coupling efficiencies of a single-user cascade are optimized in~\cite{Feng2026Coupling} and the benefit of joint placement and coupling design is demonstrated numerically. A coupled-mode cascade model is adopted in~\cite{Luo2026CMT}, where transmit power is minimized using a common coupling length. 
Other reconfigurable-antenna families reshape the channel in different ways.
Reconfigurable intelligent surfaces adjust the phases of passive reflecting elements~\cite{Wu2019}, fluid antennas switch among discrete ports~\cite{Wong2021}, and movable antennas optimize continuous positions in a region~\cite{Zhu2024}. None of these architectures has the cascaded-feed property that power extracted by one element is no longer available to downstream elements.

The shared budget considered here is the underlay cognitive-radio~(CR)~\cite{Mitola1999} constraint. Secondary users~(SUs) may transmit concurrently with primary users as long as the aggregate interference at the primary receiver~(PR) remains below an interference-temperature threshold~$\Ith$~\cite{Haykin2005, Goldsmith2009}. Capacity and power allocation under this constraint have been studied extensively for fixed antenna geometries~\cite{Ghasemi2007, Musavian2009, Kang2009}. 

Recent studies examined PASS as a secondary underlay network. In~\cite{Sun2025PASSCR}, an approximate closed-form expression for the average spectral efficiency under Rician fading is derived for a primary and a secondary waveguide, and the pinching beamforming and secondary transmit power are optimized sequentially under an interference-temperature limit. Multiple users are served by a multi-waveguide secondary base station under per-primary interference tolerances using weighted minimum mean-square error (WMMSE) with alternating optimization in~\cite{Li2026PASSCR}.
Both split a waveguide's power \emph{equally} among its pinches and neglect in-waveguide attenuation, so neither has a cascade or a controllable split; \cite{Vashakidze2025NOMA, Feng2026Coupling, Luo2026CMT} model the cascade but no shared constraint (Table~\ref{tab:related}). What is new in this paper is not the coverage but the result: under a shared budget, placement and radiation efficiency are substitutes, and the budget causes this, not the cascade. 

\begin{table}[!t]
\renewcommand{\arraystretch}{1.3}
\caption{Closest PASS Studies}
\label{tab:related}
\centering
\footnotesize
\begin{tabular}{lcccc}
\hline
 & \textbf{Cascade}   & \textbf{Power}    & \textbf{Shared}   & \textbf{Reception} \\
 & \textbf{with}      & \textbf{split a}  & \textbf{interf.}  & \textbf{model} \\
 & \textbf{atten.}    & \textbf{variable} & \textbf{budget}   & \\
\hline
\cite{Sun2025PASSCR}    & no  & no  & yes & coherent \\
\cite{Li2026PASSCR}     & no  & no  & yes & coherent \\
\cite{Vashakidze2025NOMA} & yes & yes$^{a}$ & no & coherent \\
\cite{Feng2026Coupling} & yes & yes & no  & coherent \\
\cite{Luo2026CMT}       & yes & partly$^{b}$ & no  & coherent \\
This work               & yes & yes & yes & power sum \\
\hline
\multicolumn{5}{@{}l}{\scriptsize $^{a}$Optimized under a total-power budget, not constrained to the cascade.} \\
\multicolumn{5}{@{}l}{\scriptsize $^{b}$One coupling length common to all pinches.} \\
\end{tabular}
\end{table}

We use the CR constraint as one instance of a broader class. The results depend on three of its properties: the budget is \emph{shared} by all pinches, \emph{which constraint applies is not known} when the pinches are set, and the objective \emph{does not depend} on which one applies. An exposure limit around a person, or a cap on the power leaked to an eavesdropper, has the same properties when the location is unknown. What carries over is this structure, not the magnitudes (Section~\ref{subsec:adaptivity}).

\emph{Scope.} 
We model the powers contributed by the pinches as adding \emph{incoherently} rather than as a phase-coherent beamforming sum (both PASS--CR studies of Table~\ref{tab:related} are phase-coherent). We measured what this costs and found coherence is worth $0.86\,\mathrm{bit/s/Hz}$ in our setting, but only with millimeter placement and sub-percent permittivity calibration (Remark~\ref{rem:incoherent}).
We assume neither tolerance, and so do not model phases. The resulting power-sum model lies between two coherent systems: it overstates one whose phases are uncontrolled and understates a phase-optimized one (by $0.86\,\mathrm{bit/s/Hz}$).
Every scheme is evaluated under this model, so the differences between schemes that carry the findings are differences of configuration, not of model. 
Under a phase-coherent model the magnitudes change, but the DoF remain strong substitutes and the substitutability still vanishes when the constraint is relaxed (Section~\ref{subsec:dof}).

Two properties make the problem nontrivial. First, the receiver to be protected changes, and a configuration chosen without knowing which PR is active must remain feasible for every PR it might meet, while one that knows it need only respect the constraint for that PR. The gap between the two turns out to be the largest term in the problem. Second, the position of a pinch enters twice: moving it toward its user shortens the air path but lengthens the cable path, so the signal gain is not concave in position, and the same position sets the interference at the PR, so the constraint is not convex in it either. Hence, convex programming does not apply to the joint problem.
Non-convexity does not make the problem unsolvable. A multi-start numerical search can find near-optimal solutions, and we use one as our reference. Therefore, our focus is not on another solver but on what each DoF, and knowledge of the PR location, is worth, and on how much of the rate inexpensive schemes can reach. We address this by characterizing a range of schemes, including a learned controller based on soft actor--critic (SAC).

Our contributions are as follows:

\begin{enumerate}

    \item \textbf{System model and single-link analysis.}
    We consider a CR network in which a cognitive base station~(CBS) serves $K$ SUs through $K$ repositionable pinches, under a single-feed cascade power model and an interference-temperature constraint, with a PR whose location varies. We characterize the cable--air trade-off analytically. We give the unconstrained single-link optimum for any path-loss exponent, generalizing the free-space result of~\cite{Xu2026Atten}, together with the condition that determines whether it lies on the cable at all. Restoring the constraint splits the problem into two regimes, and in the binding one the cable attenuation \emph{cancels identically}, so position trades user proximity against PR proximity instead. The model fixes the number of pinches at $K$; an efficiency of zero switches a pinch off. Nonetheless, the joint reference feeds all four pinches on $797$ of the $800$ episodes.

    \item \textbf{Problem formulation.}
    We formulate joint optimization of pinch positions $\{p_k\}$ and radiation efficiencies $\{\eta_k\}$ as a non-convex sum-rate maximization under an average interference constraint $\Ith$. 
We identify three sources of non-convexity and show that the two involving the efficiencies are removed by a change of variables to extraction shares. The allocation at fixed positions is then a convex program, which every scheme with a per-pinch allocation solves to a first-order optimality gap below $10^{-9}$. We prove that one-pass decoupled optimization can be strictly suboptimal, and separate the per-realization problem from the hedged problem that a design without the knowledge of PR location must solve.

    \item \textbf{Numerical characterization.}
    We construct a grid of schemes, three position rules by two efficiency rules. The four corners of this grid form a controlled experiment since adjacent corners differ in exactly one DoF. Therefore, differences between these schemes isolate what that DoF is worth. Since it is possible to go from the corner with no DoF (EU) to the corner with both (OP) through two paths, we measure both, and report the order-free (Shapley) attribution rather than a single path. 
This yields three quantitative design findings: 
\finding{find:subst}: Neither DoF dominates (positioning is worth $1.34\times$ radiation efficiency) and they are near-perfect substitutes.
\finding{find:constraint}: The reason for this substitutability is the interference constraint. Substitutability dissolves when the constraint is relaxed (between $\Ith = -25$ and $-15$\,dBm).
\finding{find:adapt}: PR location information matters more than optimizing both DoF together. Without it, a scheme must stay feasible for every PR location, including the one that would receive the most interference; it then radiates only $0.36\%$ of the injected power, against $28\%$ when the location is known.

\end{enumerate}

The remainder of this paper is organized as follows.
In Section~\ref{sec:system_model}, we present the system model and the
single-link analysis of the cable--air trade-off. 
In Section~\ref{sec:problem}, we formulate the joint problem, analyze its structure, and cast it as a Markov decision process (MDP) for the learned controller.
In Section~\ref{sec:results}, we characterize the problem numerically through specifically chosen schemes and establish our three findings, and in
Section~\ref{sec:conclusion}, we present our conclusions.

%% file: body_section_II_system_model.tex
\section{System Model}
\label{sec:system_model}

\subsection{Network Topology}
\label{subsec:topology}

We consider an underlay CR network with a cognitive base station (CBS), feeding a dielectric cable of length $L$ mounted horizontally at height $H$. $K$ pinches are activated along the cable at ordered positions $p_1 < p_2 < \cdots < p_K$, $p_k \in [0, L]$, jointly serving $K$ secondary users (SUs). Every pinch contributes to every user (Section~\ref{subsec:sinr}). 
Each pinch is a low-gain radiating element with radiation efficiency $\eta_k \in [0, 1]$, the ratio of the power it radiates into the air channel to the power arriving from the cable. The number of pinches is fixed at $K$, but any pinch can be switched off by setting $\eta_k = 0$. 

\begin{figure}[!t]
    \centering
    \includegraphics[width=\columnwidth]{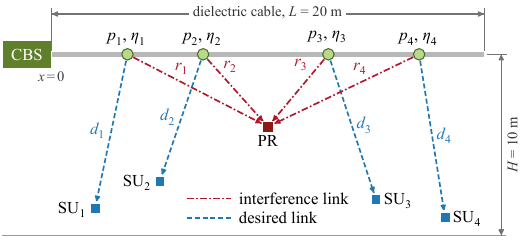}
    \caption{System model ($L = 20$\,m, $H = 10$\,m): four
        pinches at $p_1 < \cdots < p_4$ with
        efficiencies $\eta_1, \ldots, \eta_4$ serve
        SU$_1$--SU$_4$ and interfere with the
        PR. Although every pinch contributes to every user~\eqref{eq:snr}, one desired link is drawn per pinch for legibility.}
    \label{fig:system_model}
\end{figure}

We adopt a 2-D Cartesian system. The dielectric cable is along the $x$-axis at $y = H$ with the CBS feeding it at $(0, H)$. SU $k$ is located at $(x_k, y_k)$, and the PR is at $(x_p, y_p)$ as depicted in Fig.~\ref{fig:system_model}. SUs are indexed in ascending order of $x_k$. Adjacent pinches are at least $\Delta_{\min}$ apart. This is not a positioning resolution but an assumption that limits mutual coupling and keeps the model valid. 
The link distances are
\begin{align}
    d_k(p_k) &= \sqrt{(p_k - x_k)^2 + (H - y_k)^2},
    \label{eq:dk} \\
    r_k(p_k) &= \sqrt{(p_k - x_p)^2 + (H - y_p)^2}.
    \label{eq:rk}
\end{align}

The restriction to two dimensions loses no generality for isotropic elements: for a node at $(x, w, z)$ the pinch-to-node distance obeys $d^2 = (p - x)^2 + w^2 + (H - z)^2$, so a 3-D deployment maps onto the 2-D one under the remap $y_{\mathrm{eff}} = H - \sqrt{(H-z)^2 + w^2}$.

SUs are served sequentially in a TDMA architecture with equal slot allocation. In slot $k$, the CBS transmits SU $k$'s signal over the full band $B$. All $K$ pinches radiate simultaneously per the cascade model of Section~\ref{subsec:cable}. TDMA eliminates inter-user interference by construction. 

We consider two timescales. At the slow scale, time is divided into episodes. Within an episode, the SU positions and the location of the active PR are fixed. The PR to be protected may change from one deployment to the next. To account for this, we draw the PR location uniformly over a known region at the start of each episode. 
At the fast scale, an episode consists of $T$ reconfiguration intervals, each corresponding to one
fading coherence period. The small-scale fading is independent across intervals. At the start of every interval the CBS may update the configuration, moving each pinch by at most $\Delta p_{\max}$. 

\subsection{In-Cable Propagation and Power Cascade}
\label{subsec:cable}

The CBS feeds SU $k$'s signal into the cable 
with power $P_{\mathrm{total}}$. Propagation over a segment of length $\ell$ incurs attenuation and phase shift corresponding to multiplying the field by $a(\ell) = e^{-(\alphac + j\beta_c)\ell}$~\cite{Pozar2011}, with attenuation constant $\alphac \ge 0$\,Np/m. We take $\alphac = 0.05$\,Np/m ($0.43$\,dB/m) at $2.4$\,GHz, within the range of $0.08$--$1.3$\,dB/m assumed in the PASS literature~\cite{Xu2026Atten, Zhang2026Directional}; Section~\ref{subsec:dof} checks $0.08$\,dB/m. Each pinch radiates a fraction $\eta_k$ of the power arriving at it, so the arriving and radiated powers at pinch $k$ are
\begin{align}
    P_{\mathrm{arrive},k}
    &= P_{\mathrm{total}}\, e^{-2\alphac p_k}
      \prod_{j=1}^{k-1}(1 - \eta_j),
    \label{eq:power_arrive} \\
    P_{\mathrm{rad},k}
    &= \eta_k\, P_{\mathrm{total}}\, e^{-2\alphac p_k}
      \prod_{j=1}^{k-1}(1 - \eta_j).
    \label{eq:power_rad}
\end{align}
This cascade dependency distinguishes this feed from free-space movable-antenna systems, and a joint optimization of $\{p_k\}$ and $\{\eta_k\}$ is required.

\subsection{Air Channel Model}
\label{subsec:air_channel}

We model the channel from a pinch to an SU with Rician fading, appropriate for indoor LOS~\cite{3GPP38901}:
\begin{equation}
    h_{sk}
    = \sqrt{\tfrac{\kappa}{\kappa + 1}}\,\bar{h}_{sk}
    + \sqrt{\tfrac{1}{\kappa + 1}}\,\tilde{h}_{sk},
    \label{eq:rician_channel}
\end{equation}
where $\kappa \geq 0$ is the Rician factor, $|\bar{h}_{sk}| = 1$ is the deterministic LOS component and $\tilde{h}_{sk} \sim \CN(0, 1)$ represents diffuse scattering. With this normalization, the small-scale coefficient has unit mean power, and the composite channel $\sqrt{\Omega_0/d_k(p_k)^{\alpha}}\,h_{sk}$, used in~\eqref{eq:snr}, has mean power $\Omega_0/d_k(p_k)^{\alpha}$. Here, $\Omega_0$ denotes the reference path gain and $\alpha$ is the path-loss exponent. 
We denote the normalized coefficient for the path from pinch $m$ to SU $k$ by $h_{sk}^{(m)}$. Each coefficient follows~\eqref{eq:rician_channel}. Fading arises from moving scatterers, while the terminals are immobile. Therefore, the LOS component remains constant within an episode but the diffuse component changes between intervals. Hence, the diffuse component $\tilde{h}_{sk}^{(m)}$ is drawn independently for every path and every reconfiguration interval. The interference channel from pinch $k$ to the PR is modeled in the same way, with mean power $\Omega_0 / r_k(p_k)^{\alpha}$.

Because every pinch radiates to every SU, the desired links form a $K \times K$ set, with distances 
\begin{equation}
    d_{km}(p_m) \triangleq \sqrt{(p_m - x_k)^2 + (H - y_k)^2},
    \label{eq:dist_cross}
\end{equation}
where the distance from pinch $m$ to SU $k$ is denoted by $d_{km}$ so $d_{kk} = d_k$ of~\eqref{eq:dk}. The air-only gains are
\begin{align}
    \tilde{G}_{sk}^{(m)}(p_m)
    &\triangleq \frac{\Omega_0}{d_{km}(p_m)^{\alpha}},
    \label{eq:air_gain_desired} \\
    \tilde{G}_{pk}(p_k)
    &\triangleq \frac{\Omega_0}{r_k(p_k)^{\alpha}}.
    \label{eq:air_gain_interference}
\end{align}

\subsection{Cable--Air Trade-off and Optimal Pinch Position}
\label{subsec:tradeoff}

Consider a single-link system. The cascade product in~\eqref{eq:power_rad} is absent and signal power $P_{\mathrm{rad},k}\,\tilde{G}_{sk}^{(k)}(p_k)$ depends on $p_k$ only through $f_k(p_k) \triangleq e^{-2\alphac p_k} / d_k(p_k)^{\alpha}$. Moving toward the user trades cable attenuation against free-space distance, and whether an interior stationary point exists is decided by one comparison. First, we show the trade-off with the interference constraint excluded in Proposition~\ref{prop:optimal_pk}. Then we state it with the constraint in Proposition~\ref{prop:constrained_pk}.

\begin{proposition}[Unconstrained single-link pinch position]
\label{prop:optimal_pk}
Consider one pinch at $p_k \in [0, L]$ and its link to SU~$k$, where the dependence on $p_k$ is captured completely by $f_k$. Assume the user lies on the served side of the feed and within the cable's span, $0 < x_k \le L$, and define $h \triangleq H - y_k > 0$ and $u \triangleq x_k - p_k$. Then
\begin{equation}
    \frac{\mathrm{d}\ln f_k}{\mathrm{d}p_k}
    = \frac{-2\alphac\,\bigl(u^{2} - 2cu + h^{2}\bigr)}{u^{2} + h^{2}},
    \qquad c \triangleq \frac{\alpha}{4\alphac},
    \label{eq:optimality_condition}
\end{equation}
for $\alphac > 0$, and any interior stationary point satisfies $u^{2} - 2cu + h^{2} = 0$. The three possibilities are:
\begin{enumerate}
\item[(i)] \emph{Lossless:} If $\alphac = 0$ there is no trade-off: $f_k = \Omega_0^{-1}\tilde{G}_{sk}^{(k)}$ is strictly increasing on $[0, x_k)$ and strictly decreasing on $(x_k, L]$, so $p_k^\ast = x_k$.
\item[(ii)] \emph{Attenuation-dominated ($c \le h$):} The quadratic is non-negative for every $u$ (zero only at the double root $u = c$ when $c = h$, an inflection), so $f_k$ is strictly decreasing on $[0, L]$ and $p_k^\ast = 0$: the pinch is at the cable feed.
\item[(iii)] \emph{Two-root regime ($c > h$):} The quadratic has roots $u_\pm = c \pm \sqrt{c^{2} - h^{2}}$, both strictly positive. A root corresponds to a position on the cable only if $u < x_k$. Then, $x_k - u_-$ is the unique interior local maximum of $f_k$ and $x_k - u_+$ is its unique interior local minimum.
\end{enumerate}
In all three regimes $f_k$ is strictly decreasing for $p_k > x_k$, so $p_k^\ast \le x_k \le L$. For $\alpha_c > 0$, the right endpoint is never the maximizer. Therefore, in regime~(iii) the global maximum is attained either at $p_k = 0$ or, if $u_- < x_k$, at $p_k = x_k - u_-$, whichever gives the larger value of $f_k$.
\end{proposition}

\begin{proof}
For $\alphac > 0$, $\ln f_k = -2\alphac p_k - \tfrac{\alpha}{2}\ln(u^{2} + h^{2})$ gives~\eqref{eq:optimality_condition} directly, so the sign of the derivative is that of $-(u^{2} - 2cu + h^{2})$, with discriminant $4(c^{2}-h^{2})$. For $c \le h$ the quadratic is non-negative with at most one zero, so $f_k$ is strictly decreasing. This is~(ii). For $c > h$ it is negative exactly on $(u_-, u_+)$, and both roots are positive. So as $p_k$ increases, the derivative is first negative, then positive, then negative. Consequently, $x_k - u_-$ is the local maximum and $x_k - u_+$ is the local minimum. Each lies in $(0, L)$ iff its corresponding root satisfies $u < x_k$, giving us~(iii). For $p_k > x_k$, we have $u < 0$, so the quadratic is positive and $f_k$ is strictly decreasing. Thus, when $\alphac>0$, the right endpoint ($p_k=L$) cannot maximize $f_k$. Minimizing the pinch-to-user distance yields~(i).
\end{proof}

\begin{remark}
\label{rem:prop1_scope}
Regime~(iii) always applies for the parameters of Section~\ref{sec:results} ($\alpha = 2.2$, $\alphac = 0.05$, so $c = 11$\,m, $h \le 9.5$\,m). However, existence of two real roots does not guarantee an interior global maximum on the cable: First, for a user whose cable projection is near the feed, the candidate position $p_k=x_k-u_-$ may lie outside the cable. Second, it may yield a smaller value of $f_k$ than the $f_k$ at the feed ($p_k=0$). For these two cases, the optimal position is $p_k^\ast=0$.
For free-space path loss ($\alpha = 2$), the interior optimum and regime~(ii) coincide with Lemma~1 of~\cite{Xu2026Atten}; Proposition~\ref{prop:optimal_pk} also compares the feed with the interior point when both stationary points lie on the cable, which at our parameters occurs for users with $x_k > u_+ \ge 16.5$\,m.
In the lossless limit the offset $u_- = c - \sqrt{c^2 - h^2} \approx h^2/(2c) \to 0$ as $\alphac \to 0$, and Proposition~\ref{prop:optimal_pk} reduces to the single-user rule of~\cite{Ding2025Placement}, which places the pinch at the user's own cable coordinate.
The offset is the cost of cable loss: the pinch moves back toward the feed, trading a shorter cable run for a longer air path. 
\end{remark}

Proposition~\ref{prop:optimal_pk} gives the optimum $p_k$ for the \emph{unconstrained} trade-off between the cable distance and the free-space distance. 
Restoring the interference constraint does not just shift the optimum; it changes the quantity that the position balances, and the optimum can be characterized explicitly.

\begin{proposition}[Interference-constrained single-link optimum]
\label{prop:constrained_pk}
Consider the same single link, now subject to the $K = 1$ case of~\eqref{eq:avg_interference_constraint}, namely $\eta\,P_{\mathrm{total}}\,\Omega_0\,e^{-2\alphac p}/r(p)^{\alpha} \le \Ith$, over $\eta \in [0,1]$ and $p \in [0,L]$. We define $A(p) \triangleq e^{-2\alphac p}/d(p)^{\alpha}$ and $B(p) \triangleq e^{-2\alphac p}/r(p)^{\alpha}$. The optimal efficiency at a given position is $\eta^\ast(p) = \min\{1,\ \Ith/(P_{\mathrm{total}}\Omega_0 B(p))\}$, and the corresponding received signal power is
\begin{equation}
    S^\ast(p)
    = \min\Bigl\{\,
        P_{\mathrm{total}}\Omega_0 A(p),
        \;\;
        \Ith \bigl(r(p)/d(p)\bigr)^{\alpha}
      \Bigr\},
    \label{eq:two_regimes}
\end{equation}
so that $p^\ast$ maximizes the pointwise minimum of two functions of position. The two branches are two regimes:
\begin{enumerate}
\item[(i)] \emph{Slack.} Where the first branch is the smaller, the constraint is inactive, $\eta^\ast = 1$, and the position problem is that of Proposition~\ref{prop:optimal_pk}: cable attenuation is traded against free-space distance and the optimum is pulled back from the user by $u_-$.
\item[(ii)] \emph{Binding.} Where the second branch is the smaller, $\eta^\ast < 1$ and \emph{the cable attenuation cancels identically}. $S^\ast$ then depends on $p$ only through the geometric ratio $r(p)/d(p)$: the optimal position maximizes the PR-to-user distance ratio and is independent of $\alphac$, $P_{\mathrm{total}}$ and $\Ith$, which only scales $S^\ast$.
\end{enumerate}
\end{proposition}

\begin{proof}
The rate is strictly increasing in $S = \eta P_{\mathrm{total}}\Omega_0 A(p)$, so the constraint is tight at the optimum unless $\eta = 1$ is feasible, which gives $\eta^\ast(p)$. Substituting it gives~\eqref{eq:two_regimes}. For the second argument, $A/B = (r(p)/d(p))^{\alpha}$ because $e^{-2\alphac p}$ is common to $A$ and $B$.
\end{proof}

\begin{remark}
\label{rem:two_regimes}
When the budget limits the radiated power, cable loss attenuates the desired signal and the interference alike, 
so the best position does not depend on it. That is, the choice of position is no longer about balancing cable loss against distance to the user, but about balancing the distance to the PR against the distance to the user.
With per-pinch efficiencies the cancellation carries over to $K$ pinches: written in the positions and the radiated powers, \eqref{eq:avg_interference_constraint} and every $\SNR_k$ are free of $\alphac$, which then enters only through the total-power limit $\sum_k P_{\mathrm{rad},k}\,e^{2\alphac p_k} \le P_{\mathrm{total}}$. Lowering $\alphac$ only loosens that limit, so where it is slack the optimal rate does not change.
\end{remark}

\begin{remark}
\label{rem:multiuser}
Propositions~\ref{prop:optimal_pk} and~\ref{prop:constrained_pk} are single-user statements. With multiple users three couplings arise. First, the shared budget for $\Ith$ means that moving pinch $k$ consumes budget that narrows every other pinch's feasible range. Second, the cascade product $\prod_{j<k}(1-\eta_j)$ means the incoming power at pinch $k$ depends on all upstream efficiencies. Third, every pinch contributes to \emph{every} user's received power in~\eqref{eq:snr}, so moving $p_k$ changes \emph{all} rates. Hence, positions and efficiencies have to be chosen jointly, and the single-user optima of Propositions~\ref{prop:optimal_pk} and~\ref{prop:constrained_pk} need not hold for a multi-user case.
\end{remark}

\subsection{Received Signal Model and SNR}
\label{subsec:sinr}

In slot $k$, all $K$ pinches radiate SU~$k$'s waveform simultaneously, so SU~$k$ receives it over $K$ paths, each with its own phase. The phases are not controlled, and we model the received power as the sum of the path powers (Remark~\ref{rem:incoherent}). The SNR at SU~$k$ is then
\begin{equation}
	\SNR_k
	= \frac{\sum_{m=1}^{K} P_{\mathrm{rad},m}\,\tilde{G}_{sk}^{(m)}(p_m)\,|h_{sk}^{(m)}|^{2}}
	{N_0 B},
	\label{eq:snr}
\end{equation}
with $N_0 B$ the noise power. The TDMA time-averaged spectral efficiency of SU~$k$ is
\begin{equation}
    R_k = \frac{1}{K}\log_2\!\left(1 + \SNR_k\right),
    \label{eq:rate}
\end{equation}
where the factor $1/K$ is the TDMA duty cycle. The total secondary sum spectral efficiency, in bit/s/Hz, is
\begin{equation}
    \Rtotal(\vp, \veta) = \sum_{k=1}^{K} R_k,
    \label{eq:sum_rate}
\end{equation}
and the rate of interest is its average over the fading, $\bar{R}_{\mathrm{total}}(\vp,\veta) \triangleq \Ex_{h}\bigl[\Rtotal(\vp,\veta)\bigr]$. In Section~\ref{subsec:opt_problem}, we present the approximation maximized by the schemes that do not observe the fading.

\begin{remark}
\label{rem:incoherent}
Adding the powers of the $K$ paths rather than their amplitudes makes the mean received power at SU $k$ equal to $\sum_{m=1}^{K} P_{\mathrm{rad},m}\,\tilde{G}_{sk}^{(m)}(p_m)$. 
This holds on average. Per-path phases are $\theta_{km} = \beta_c p_m + k_0 d_{km}$, where $\beta_c$ is the propagation constant in the cable and $k_0$ is the free-space wavenumber. When the coherent received power is averaged over uniformly distributed phases on $[0,2\pi)$, cross terms drop and what is left is the incoherent power sum.
By Jensen's inequality,~\eqref{eq:rate} is an \emph{upper} bound on what a system with uncontrolled phases achieves. 
Controlling the phases is demanding. 
Along the cable, each traveled wavelength rotates the phase by $2\pi$. Since this wavelength is small, $\lambda_c = \lambda/\sqrt{\varepsilon_{\mathrm{eff}}} \approx 8.3$\,cm for the $\varepsilon_{\mathrm{eff}} = 2.25$ we assume, where $\lambda = 12.5$\,cm is the free-space wavelength at $2.4$\,GHz, a small change in pinch position may turn a constructive combination into a cancellation. 
Furthermore, an error in the permittivity $\varepsilon_{\mathrm{eff}}$ shifts every phase along the cable. We measured both effects. On $60$ random
episodes of Section~\ref{sec:results}, a phase-aware design (positions optimized, one common efficiency) scores $0.86\,\mathrm{bit/s/Hz}$ above its incoherent counterpart on the fading approximation~\eqref{eq:surrogate}. In a separate test on $30$ episodes, a placement error of $4$\,mm already costs $11\%$ of the gain measured there
and at $16$\,mm about $90\%$ is gone.
Separately, a permittivity error of $0.2\%$ costs $10\%$ of the gain, and an error of $2\%$ removes it entirely.
Therefore, we assume neither millimeter placement nor sub-percent calibration, and instead check Findings~\ref{find:subst} and~\ref{find:constraint} under a coherent model in Section~\ref{subsec:dof}.

\end{remark}

\subsection{CSI Model and Interference Constraint}
\label{subsec:csi}

CSI is \emph{partial}. For the desired links the CBS estimates the full $K \times K$ matrix of per-path gains  $|h_{sk}^{(m)}|^{2}$ from uplink pilots. This requires $K^2$ measurements per reconfiguration interval, whereas the expected-gain schemes require no instantaneous measurements and use only large-scale parameters. 
For the interference links, only the large-scale parameters $(\Omega_0, \alpha, \kappa)$ and the expected gains $\tilde{G}_{pk}(p_k) = \Omega_0/r_k(p_k)^\alpha$ of the active PR are known, and they are re-acquired each episode. The hedged schemes of Section~\ref{subsec:adaptivity} are the exception: they know only the region in which the PR may lie. Since the $K$ expected gains over-determine $(x_p, y_p)$, the PR location can be recovered from them in closed form, so the schemes that we define in terms of $(x_p, y_p)$ implicitly assume this recovery step.

Instantaneous interference CSI at the CBS is unrealistic, so we adopt an \emph{average interference constraint}. Under TDMA all $K$ pinches radiate in every slot and, by the incoherent addition of Remark~\ref{rem:incoherent}, the expected interference is the same in every slot, so the frame-averaged constraint reduces to
\begin{multline}
    \sum_{k=1}^{K}
    P_{\mathrm{rad},k}\, \tilde{G}_{pk}(p_k) \\
    = \sum_{k=1}^{K}
      \frac{\eta_k\, P_{\mathrm{total}}\,\Omega_0\,
            e^{-2\alphac p_k}}{r_k(p_k)^{\alpha}}
      \prod_{j=1}^{k-1}(1-\eta_j)
    \leq \Ith.
    \label{eq:avg_interference_constraint}
\end{multline}
This formulation follows~\cite{Goldsmith2009} and is the usual one for underlay CR without instantaneous PR feedback~\cite{Ghasemi2007, Musavian2009}. It couples all $\{p_k, \eta_k\}$: each position sets both an interference gain and a cable loss, and each $\eta_j$ changes, through the cascade product, the radiated power and hence the interference of every downstream pinch.

%% file: body_section_III_problem_formulation.tex
\section{Problem Formulation}
\label{sec:problem}

\subsection{Optimization Problem}
\label{subsec:opt_problem}

The optimization variables are the vector of pinch positions $\vp = [p_1,\ldots,p_K]^T$ and the vector of radiation efficiencies $\veta = [\eta_1,\ldots,\eta_K]^T \in [0,1]^K$, and we call a pair $(\vp, \veta)$ a configuration. Together, they form a continuous $2K$-dimensional space. Total transmit power $P_{\mathrm{total}}$ is a fixed parameter: the CBS always operates at full power. The joint sum-rate maximization problem is
\begin{align}
    \mathbf{(P1):}\quad
    \max_{\vp,\, \veta} \quad
    & \bar{R}_{\mathrm{total}}(\vp, \veta)
    \label{eq:P1} \\[4pt]
    \text{s.t.} \quad
    & \textstyle\sum_{k=1}^{K} P_{\mathrm{rad},k}\,\tilde{G}_{pk}(p_k) \leq \Ith,
    \tag{C1}\label{eq:C1} \\[4pt]
    & 0 \leq \eta_k \leq 1,
      \quad k = 1, \ldots, K,
    \tag{C2}\label{eq:C2} \\[2pt]
    & 0 \leq p_1, \quad p_K \leq L, \nonumber \\
    & p_k \geq p_{k-1} + \Delta_{\min},
      \quad k = 2, \ldots, K,
    \tag{C3}\label{eq:C3} \\[2pt]
    & |p_k^{t+1} - p_k^{t}| \leq \Delta p_{\max},
      \quad \forall k,
    \tag{C4}\label{eq:C4}
\end{align}
where $\bar{R}_{\mathrm{total}}$ is the fading-averaged sum rate of Section~\ref{subsec:sinr} and $P_{\mathrm{rad},k}$ is given by~\eqref{eq:power_rad}.
Constraints~\eqref{eq:C1}--\eqref{eq:C4} are for the average interference temperature limit, physical efficiency bounds, cable ordering with a minimum separation between adjacent pinches, and maximum per-step displacement, where $t$ indexes the reconfiguration intervals at which the CBS may adjust the pinch positions (formalized in Section~\ref{subsec:mdp}). 
Constraint~\eqref{eq:C4} applies only to the learned controller of Section~\ref{subsec:mdp}, since it moves the pinches incrementally. Other schemes presented in Section~\ref{sec:results} fix their positions once per episode; thus, \eqref{eq:C4} does not apply to them. Since $T\,\Delta p_{\max} = 40\,\mathrm{m}$ exceeds $L$, every fixed configuration is reachable by the controller within half an episode.

The fading average in~\eqref{eq:P1} has no closed form. Therefore, among the schemes reported in Section~\ref{sec:results}, those that do not have instantaneous fading information maximize its second-order approximation,
\begin{equation}
    \frac{1}{K}\sum_{k=1}^{K}\Bigl[\log_2(1+\mu_k)
    - \frac{c_\kappa \sum_{m=1}^{K} s_{km}^{2}}{2\ln 2\,(1+\mu_k)^{2}}\Bigr],
    \label{eq:surrogate}
\end{equation}
where $s_{km} \triangleq P_{\mathrm{rad},m}\,\tilde{G}_{sk}^{(m)}(p_m)/(N_0 B)$ is the average SNR that pinch~$m$ contributes to SU~$k$, $\mu_k \triangleq \sum_{m} s_{km}$, and $c_\kappa \triangleq (1+2\kappa)/(1+\kappa)^2$ is the relative variance of the Rician power gain. 
The rates reported in Section~\ref{sec:results} are averages of~\eqref{eq:sum_rate} over the $T$ reconfiguration intervals of an episode, each with independent fading.

Constraint~\eqref{eq:C3} requires adjacent pinches to be separated by at least $\Delta_{\min}$ and ensures that all pinch positions are within the cable length. For fair comparison, this requirement is applied to every scheme in Section~\ref{sec:results}. We allow $p_1=0$, i.e., the minimum separation does not apply between the feed and the first pinch. The joint reference scheme exploits this by placing the first pinch at or near the feed in a substantial fraction of episodes. The feasible position set is
\begin{equation}
    \mathcal{F} = \bigl\{\vp:\; 0 \leq p_1,\;
        p_k \geq p_{k-1} + \Delta_{\min},\;
        p_K \leq L \bigr\}.
    \label{eq:clip}
\end{equation}
The same set is used by every scheme. The learned controller's moves are projected onto $\mathcal{F}$ exactly, with the pool-adjacent-violators algorithm; in the shifted coordinates $p_k - k\Delta_{\min}$, the separation constraints become monotonicity constraints.

\subsection{Per-Realization and Hedged Formulations}
\label{subsec:regimes}

Since the PR location varies between episodes, \eqref{eq:P1} can be approached in two ways. One may solve~\eqref{eq:P1} by re-acquiring the PR location at every episode. This is the \emph{per-realization} problem, whose value $\bar{R}^\star_{\mathrm{total}}(x_p, y_p)$ is attainable only by a scheme that knows where the PR is. Alternatively, for the current SU layout, one may seek a single configuration that is feasible for every PR location the region allows. This is the \emph{hedged} problem,
\begin{equation}
    \max_{\vp,\,\veta}\ \bar{R}_{\mathrm{total}}(\vp,\veta)
    \quad \text{s.t.} \quad
    \sup_{(x_p,y_p)\in\mathcal{P}} I(\vp,\veta;x_p,y_p) \leq \Ith,
    \label{eq:P1static}
\end{equation}
with $\mathcal{P}$ the region where the PR may be.
Since the objective does not depend on the PR --- the PR enters only through~\eqref{eq:C1} --- the difference between $\Ex[\bar{R}^\star_{\mathrm{total}}]$ and the value of~\eqref{eq:P1static} 
is, by definition, the price of hedging against the worst-case PR location. We measure it in Section~\ref{subsec:adaptivity} and it is large.
One point worth noting is that the solver approximates the supremum by 
enforcing~\eqref{eq:C1} on $241$ PR positions that are $7.5$\,cm apart along the region's edge nearest the cable, $y_p = 9.5$\,m, where the interference of any configuration is largest. Increasing that resolution for the solver becomes computationally expensive. However, a finer resolution check only for the interference is practical and 
reveals violations of $\Ith$ of up to $0.61\%$ ($0.58\%$ on average). Removing them would reduce OP-hedged's rate by $0.008\,\mathrm{bit/s/Hz}$ on average, so the price of hedging is, if anything, slightly understated. Feasibility is therefore guaranteed at the $7.5$\,cm resolution, not over the whole continuous region.

\subsection{Non-Convexity Analysis}
\label{subsec:nonconvexity}

\label{subsubsec:qvars}
TDMA removes inter-user interference from $\SNR_k$, but problem~\eqref{eq:P1} is non-convex in three ways. 
(i)~The power that pinch~$k$ delivers to SU~$k$ is proportional to $f_k(p_k) = e^{-2\alphac p_k}/d_k(p_k)^\alpha$. At the considered parameters, it falls in regime~(iii) of Proposition~\ref{prop:optimal_pk} and is not concave on $[0, L]$. Since all positions must satisfy~\eqref{eq:C1}, none can be chosen independent of the others. 
(ii)~For fixed $\vp$ the cascade product $\prod_{j<k}(1-\eta_j)$ makes the objective non-separable in $\veta$. 
(iii)~The constraint~\eqref{eq:C1} is non-convex in both sets of variables. In $\veta$, the interference is linear in each efficiency taken alone, but contains products of the efficiencies of different pinches. Such a function is neither convex nor concave, and the set of $\veta$ on which it stays below $\Ith$ can be non-convex.
In $\vp$, interference contains the factor $e^{-2\alphac p_k}/r_k(p_k)^\alpha$. It rises as pinch~$k$ approaches the PR and then falls once it has passed it. Therefore, a pinch can satisfy~\eqref{eq:C1} on either side of the PR but not close to it, so the positions allowed by~\eqref{eq:C1} can form two separate intervals.

Sources~(ii) and~(iii) in $\veta$ arise from the choice of variables. Let $q_k \triangleq \eta_k \prod_{j<k}(1-\eta_j)$ be the share of the injected power that pinch $k$ extracts. 
Expressed this way, the cascade reduces to a single condition $\vect{q} \ge 0$, $\sum_k q_k \le 1$,
since $\sum_k q_k = 1 - \prod_k(1-\eta_k)$ 
and every such $\vect{q}$ corresponds to an $\veta$, through $\eta_k = q_k/(1 - \sum_{j<k} q_j)$.

In these variables $P_{\mathrm{rad},k} = q_k\,P_{\mathrm{total}}\,e^{-2\alphac p_k}$ is linear, so at fixed $\vp$ the interference in~\eqref{eq:C1} is a single linear inequality, the constraint set is a polytope, and every $\SNR_k$ of~\eqref{eq:snr} is affine in $\vect{q}$. Therefore, the objective of~\eqref{eq:P1} is concave, and the allocation at fixed positions is a convex program. Source~(i) and the dependence of the constraint coefficients on $\vp$ remain non-convex. 
Thus, for the schemes in Section~\ref{sec:results}, every fixed-position allocation is solved to a first-order (Frank--Wolfe) optimality gap below $10^{-9}$. The gap bounds the suboptimality where the objective is concave; the schemes maximize the approximation~\eqref{eq:surrogate}, whose concavity is checked rather than proven (Section~\ref{subsubsec:etarule}).

\subsection{Suboptimality of One-Pass Decoupling}
\label{subsec:decoupled}

Throughout this subsection the rate is evaluated at the average gains. A natural heuristic decouples~\eqref{eq:P1}: choose $\vp$ by a rule that does not anticipate the allocation, then choose $\veta$ for that $\vp$. 
Accordingly, we consider a one-pass decoupling, and show that fixing the positions before the allocation has its cost. On some geometries, a small move away from the decoupled solution raises the sum-rate in proportion to the move, so the solution is not even a local optimum.

Moving a pinch to reduce its interference gain $\tilde{G}_{pk}$ frees some of the budget in~\eqref{eq:C1}  that can be spent on a downstream efficiency. However, the decoupled procedure cannot make use of it because 
it places each pinch for its own user without taking~\eqref{eq:C1} into account. 
The moved pinch also either increases or decreases the cross-path gains in~\eqref{eq:snr}, modifying the received power of every other user. 
Consequently, the exchange improves the sum-rate only when the benefit of reallocating the released budget exceeds any loss that may be caused by the cross-path changes. We formalize this condition with the following proposition.

\begin{proposition}[Suboptimality of one-pass decoupling]
\label{prop:suboptimal}
Let $K = 2$ and consider the one-pass procedure that first places $p_k$ at the unconstrained single-link optimum of Proposition~\ref{prop:optimal_pk} for SU~$k$, and then maximizes $\Rtotal$ over $\veta$ subject to~\eqref{eq:C1}--\eqref{eq:C2}. Write $(\vp^*, \veta^*)$ for its output and assume
\begin{enumerate}
\item[(A1)] \eqref{eq:C1} is active at $(\vp^*, \veta^*)$;
\item[(A2)] $\eta_1^* > 0$ and $\eta_2^* < 1$;
\item[(A3)] $p_1^*$ is an interior stationary point of $f_1$, i.e.\ regime~(iii) of Proposition~\ref{prop:optimal_pk} with $u_- < x_1$.
\end{enumerate}
Every path out of pinch~$1$ has the form $\eta_1 P_{\mathrm{total}}\Omega_0 e^{-2\alphac p_1}/s(p_1)^{\alpha}$, so its logarithmic derivative at $p_1^*$ is
\begin{equation}
    \nu_\bullet
    \triangleq -2\alphac + \frac{\alpha\,(\xi_\bullet - p_1^*)}{s_\bullet^{2}},
    \label{eq:logderiv}
\end{equation}
with $(\xi_\bullet, s_\bullet)$ equal to $(x_1, d_{11})$ for the own path~$\nu_1$, $(x_2, d_{21})$ for the cross path~$\nu_2$, and $(x_p, r_1)$ for the interference path~$\nu_p$, all evaluated at $p_1^*$; note $\nu_1 = 0$ by~(A3) and Proposition~\ref{prop:optimal_pk}. If the geometry satisfies
\begin{equation}
    \nu_p > 0
    \quad\text{and}\quad
    \frac{\nu_p}{r_1^{\alpha}}
    \left(\frac{r_2}{d_{22}}\right)^{\!\alpha}
    > \frac{\nu_2}{d_{21}^{\alpha}},
    \tag{G}
    \label{eq:condG}
\end{equation}
where $r_2$ and $d_{22}$ are taken at $p_2^*$, then $(\vp^*, \veta^*)$ is not a global optimum of~\eqref{eq:P1}, so $\Rtotal(\vp^*, \veta^*) < \Rtotal(\vp^\dagger, \veta^\dagger)$ for every global optimum $(\vp^\dagger, \veta^\dagger)$. Geometries satisfying~\eqref{eq:condG} exist at the parameters of Section~\ref{sec:results}.
\end{proposition}

\begin{proof}
Define $c_k(p) \triangleq P_{\mathrm{total}}\Omega_0 e^{-2\alphac p}/r_k(p)^{\alpha} > 0$. Now~\eqref{eq:C1} can be written as $\eta_1 c_1(p_1) + \eta_2(1-\eta_1)c_2(p_2) \le \Ith$. 
Let $p_1 = p_1^* - \epsilon$, with $\epsilon > 0$, i.e. allow a small change in $p_1$, while keeping $p_2 = p_2^*$ and $\eta_1 = \eta_1^*$. Let $\eta_2$ take the value that keeps~\eqref{eq:C1} tight,
\begin{equation*}
    \eta_2(\epsilon)
    = \frac{\Ith - \eta_1^* c_1(p_1^* - \epsilon)}
           {(1-\eta_1^*)\,c_2(p_2^*)}.
\end{equation*}

\emph{Feasibility.} Moving $p_1$ toward the feed increases $p_2 - p_1$, so~\eqref{eq:C3} holds, and $p_1^* - \epsilon > 0$ by~(A3); since $c_1'(p_1^*) = \nu_p\,c_1(p_1^*) > 0$, $\eta_2(\epsilon) = \eta_2^* + \Theta(\epsilon)$ is increasing and~(A2) keeps it below $1$, so~\eqref{eq:C2} holds.

\emph{First-order effect.} Only the path of the first pinch depends on $p_1$, and by~\eqref{eq:logderiv} each differentiates to $\nu_\bullet$ times itself. Raising $\eta_2$ by $\delta\eta_2$ adds $\delta\eta_2 (1-\eta_1^*)P_{\mathrm{total}}\Omega_0 e^{-2\alphac p_2^*}/d_{k2}^{\alpha}$ to SU~$k$'s received power, and $\delta\eta_2 = \epsilon\,\eta_1^* c_1(p_1^*)\nu_p / [(1-\eta_1^*)c_2(p_2^*)]$. Letting $\Sigma \triangleq \eta_1^* P_{\mathrm{total}}\Omega_0 e^{-2\alphac p_1^*}$, so that $\eta_1^* c_1(p_1^*) = \Sigma/r_1^{\alpha}$, and cancelling $P_{\mathrm{total}}\Omega_0 e^{-2\alphac p_2^*}$ against $c_2(p_2^*)$ to leave $(r_2/d_{k2})^{\alpha}$,
\begin{align*}
    \Delta S_1
    &= \epsilon\,\Sigma\,
       \frac{\nu_p}{r_1^{\alpha}}
       \left(\frac{r_2}{d_{12}}\right)^{\!\alpha}
       + O(\epsilon^2), \\
    \Delta S_2
    &= \epsilon\,\Sigma
       \left[
         \frac{\nu_p}{r_1^{\alpha}}
         \left(\frac{r_2}{d_{22}}\right)^{\!\alpha}
         - \frac{\nu_2}{d_{21}^{\alpha}}
       \right]
       + O(\epsilon^2).
\end{align*}
SU~$1$'s own path from pinch~$1$ contributes nothing at first order because $\nu_1 = 0$; 
SU~$2$'s cross path from pinch~$1$ gives the subtracted term, which \eqref{eq:C1} does not account for. Condition~\eqref{eq:condG} makes both coefficients of $\epsilon$ strictly positive, so for all sufficiently small $\epsilon$ we have $\Delta S_1 > 0$ and $\Delta S_2 > 0$. Moreover, since each $R_k$ is strictly increasing in $S_k$, $\Delta\Rtotal = \Theta(\epsilon) > 0$. Consequently, moving pinch~$1$ and raising $\eta_2$ in this way gives a feasible and strictly better configuration, hence $(\vp^*, \veta^*)$ is not a global optimum.

\emph{Existence.} Condition~\eqref{eq:condG} holds whenever $\nu_p > 0 \ge \nu_2$, and their signs depend on those nodes' positions relative to pinch~$1$. With $\alpha = 2.2$ and $\alphac = 0.05$, $\nu_\bullet \le 0$ exactly when $(\xi_\bullet - p_1^*)/s_\bullet^{2} \le 2\alphac/\alpha = 0.0455\,\mathrm{m}^{-1}$. Therefore, an SU~$2$ at a horizontal distance of $3$\,m and a drop of $9.5$\,m gives $\nu_2 < 0$, while a PR at the same horizontal distance and a drop of $2$\,m gives $\nu_p > 0$. Both are inside the deployment of Section~\ref{sec:results}. The exchange is worthwhile when the PR is close to the cable, so that moving pinch~$1$ away from it frees budget quickly, while the other user is far below the cable, so that its cross-path gain hardly changes with the position of pinch~$1$.
\end{proof}

With Proposition~\ref{prop:suboptimal} we show that the gap exists, but only for the one-pass rule. We also did not show how large it is. 
OP-seq of Table~\ref{tab:rungs} optimizes all positions jointly for a fixed $\veta$ before allocating. It is stronger and the proposition does not cover it. 
With Finding~\ref{find:subst} of Section~\ref{sec:results}, we show that the gap is bounded. Since the two DoF are near-perfect substitutes, OU alone is only $0.066\,\mathrm{bit/s/Hz}$ below OP, and OP-seq, which allocates exactly at OU's positions, is never below OU, so its deficit cannot exceed $0.066$. We measure it on the pinned $800$ episodes of Section~\ref{sec:results}.
The gap for OP-seq is $0.028\,\mathrm{bit/s/Hz}$ (paired standard error $0.002$). Doing the allocation after the positions recovers $57\%$ of the efficiency gain, and on the approximation~\eqref{eq:surrogate} OP-seq is strictly worse on $641$ of the $800$ episodes and identical on the other $159$.

Therefore, decoupling can lose rate, but very little, and the shared budget accounts for both. A joint search can find improvements that a decoupled one may miss because the constraint couples the two DoF. Furthermore, the constraint also makes them substitutes (Findings~\ref{find:subst} and~\ref{find:constraint}), so those improvements are small.

\subsection{Reformulation as a Markov Decision Process}
\label{subsec:mdp}

Sections~\ref{subsec:nonconvexity} and~\ref{subsec:decoupled} rule out two shortcuts: convex programming, since~\eqref{eq:P1} is not jointly convex, and one-pass decoupling, which can lose rate. Therefore, in Section~\ref{sec:results} we use a joint numerical search, OP, as a near-optimal per-realization reference while characterizing~\eqref{eq:P1} through a grid of schemes. Among them is a learned MDP policy that
acts on the two timescales of Section~\ref{subsec:topology}. At each reconfiguration interval $t = 1, \ldots, T$ it sets the efficiencies and moves the pinches subject to~\eqref{eq:C4},  starting each episode from random ordered positions and equal efficiencies. 
Since SU and PR locations change between episodes and the gains fluctuate within them, the controller has to learn a mapping from what it observes to a configuration, not a single configuration.

The state at step $t$ contains the expected gains of all $K^2$ pinch--SU paths of~\eqref{eq:dist_cross}, the interference-link gains, the current positions and efficiencies, the remaining interference budget $(\Ith - I)/\Ith$ (negative if exceeded), and the ratio of instantaneous to expected gain on every pinch--SU path.

The action moves each pinch by $\Delta p_k^t \in [-\Delta p_{\max}, \Delta p_{\max}]$ and sets the relative efficiencies of the pinches on a log scale, together with an overall level that may keep them below what the interference budget allows. 
The positions $\vp^t + \Delta\vp^t$ are projected onto $\mathcal{F}$ of~\eqref{eq:clip} and scaled back towards $\vp^t$ if any pinch would move by more than $\Delta p_{\max}$.
This restores~\eqref{eq:C4} and keeps feasibility since $\mathcal{F}$ is convex, and the new positions take effect at the next interval. Then, a bisection scales the efficiency weights to the largest values satisfying~\eqref{eq:C1}, with $\eta_k\le1$, after which the overall level may reduce them further. 
Although this construction does not guarantee~\eqref{eq:C1} analytically, the constraint is satisfied at all $160\,000$ evaluated decisions of the ten trained policies at the nominal threshold. The reward is $\Rtotal(\vp^t,\veta^t)$ minus a fairness term $\omega\,\mathrm{Var}(\{R_k^t\})$ and an interference-violation penalty. The penalty is zero at every evaluated decision, and Section~\ref{sec:results} reports sum-rate rather than reward.

%% file: body_section_V_simulation_results.tex
\section{Numerical Characterization}
\label{sec:results}

\subsection{Simulation Setup and Evaluation Protocol}
\label{subsec:setup}

We consider an indoor environment with $20\,\mathrm{m}$ length and $10\,\mathrm{m}$ height. The cable fed by the CBS spans the full length along the ceiling, at $H = 10\,\mathrm{m}$. 
At the start of each episode, $K = 4$ SUs are placed at random in $[1, 19]\,\mathrm{m} \times [0.5, 8]\,\mathrm{m}$ (uniformly), and the PR location is drawn from $[1,19]\,\mathrm{m} \times [6,9.5]\,\mathrm{m}$ (also uniformly). The PR region is purposefully chosen closer to the cable than the SUs, so that~\eqref{eq:C1} binds at the nominal threshold. In Section~\ref{subsec:sensitivity}, we present a sweep on how binding it is. The remaining parameters are in Table~\ref{tab:sim_params}.

\begin{table}[!t]
\renewcommand{\arraystretch}{1.3}
\caption{Simulation Parameters}
\label{tab:sim_params}
\centering
\footnotesize
\begin{tabular}{@{}llll@{}}
\hline
$L$, $H$ & $20$\,m, $10$\,m & $\kappa$ & $5$\,dB \\
$K$ & $4$ & $N_0 B$ & $-104$\,dBm \\
$(x_p, y_p)$ & $\sim\mathcal{U}$, per episode & $P_{\mathrm{total}}$ & $24$\,dBm \\
$\Delta_{\min}$ & $0.5$\,m & $\Ith$ & $-35$\,dBm \\
$\alphac$ & $0.05$\,Np/m & $\Delta p_{\max}$ & $0.2$\,m \\
$\alpha$ & $2.2$ & $B$ & $10$\,MHz \\
$\Omega_0$ & $-30$\,dB & $T$ & $200$ steps \\
\hline
\end{tabular}
\end{table}

\subsubsection{The controlled experiment}
\label{subsubsec:ladder}
Instead of proposing a single method, we first present a grid, corners of which are generated by varying the two DoF one at a time. Along the position axis ($\vp$) we have equally spaced pinches (E) (pinch $k$ sits at $kL/(K+1)$), pinches above users (A) and optimized pinch positions per episode (O). Along the radiation efficiency axis ($\veta$) we have a single uniform efficiency for all $K = 4$ pinches (U) and efficiency per pinch (P) --- resulting in the six cells EU, AU, OU, EP, AP and OP of Table~\ref{tab:rungs}. While EU and EP give us what happens when position DoF is off, AU and AP give us the value of using the position DoF in the cheapest way, with a simple positioning rule (pinches above users, while still respecting the minimum pinch separation $\Delta_{\min}$).
To determine how much each DoF contributes, we need to measure the effect of moving from one end of an axis to the other. Therefore, the four corner cells (EU, EP, OU and OP) form the controlled experiment. OP is the cell with both DoF jointly optimized under~\eqref{eq:C1}.  It is taken as the  reference for all other schemes.

\subsubsection{Off-grid schemes}
\label{subsubsec:offgrid}
We extend our analysis with six off-grid schemes and a learned controller.
Each off-grid extension scheme investigates an axis the grid lacks: 
OP-seq gives us the value of joint optimization compared to sequential optimization. It optimizes the positions first, under a uniform efficiency (OU positions). Then it solves per-pinch allocation on those positions. 
EU-hedged and OP-hedged are used to evaluate the value of the PR location information. When the solver does not know where the PR is, it has to satisfy~\eqref{eq:C1} for every possible PR location in the support. Thus it has to \emph{hedge} against uncertainty in the PR location. 
The hedged schemes are used to make comparisons between likes in Section~\ref{subsec:adaptivity}.
OP-free is OP without~\eqref{eq:C1}; it gives the cost of the constraint.
OP-inst is OP with the instantaneous fading information: it re-solves $\veta$ at OP's positions on instantaneous SU link gains, and thus at a much finer timescale (each interval), instead of each episode. It gives us the headroom of having instantaneous 
channel information compared to a scheme that uses expected gains. 
H1 uses OP's positions, then once per episode scales $K$ single-pinch vectors to meet the interference budget. A single-pinch vector is an efficiency vector with all power extracted from a single pinch. 
H1 then evaluates them on instantaneous 
channel gains at each interval and selects the best one. H1 $-$ OP gives how much of that headroom is obtained without re-solving the allocation each interval, by a simple selection rule, whereas OP-inst $-$ H1 gives the part that needs the re-solve.

\subsubsection{The learned controller}
\label{subsubsec:SAC}
It is trained with soft actor--critic (SAC)~\cite{Haarnoja2018} on the MDP of Section~\ref{subsec:mdp}. 
It updates $(\vp,\veta)$ over the $T$ reconfiguration steps of each episode. 
The learned policy satisfies~\eqref{eq:C1} at all $160\,000$ evaluated decisions of the ten trained policies at the nominal threshold, so the interference penalty in its reward is never active there. The reward also contains a fairness term (Section~\ref{subsec:mdp}), so SAC does not maximize sum rate alone. Consequently, its sum rate may understate what a policy trained for sum rate alone would achieve.

The policy is trained with ten independent seeds and $7.5\times10^{5}$ steps for each seed.
The best checkpoint is selected using a validation episode stream and evaluated on a disjoint test stream of $80$ episodes, for each seed. With one stream per seed, a fixed set of $800$ episodes is formed. 
All of the schemes are scored on this set and SAC achieves $23.53\,\mathrm{bit/s/Hz}$, placing it between AP and OU. SAC is the only scheme that does not rely on a closed-form rule or a per-episode numerical solve.

For the fixed set of $800$ test episodes mentioned above, the SU layouts, PR locations, and fading realizations are stored explicitly. Therefore, all schemes can be evaluated under identical, paired conditions. 
Training is not deterministic. Replay indices and warm-up exploration use an unseeded generator, so SAC is reported as the mean over the ten seeds. Its evaluation from the stored checkpoints is exactly reproducible.

\begin{table}[!t]
\caption{The Grid, and What Each Off-Grid Scheme Changes}
\label{tab:rungs}
\centering
\footnotesize
\setlength{\tabcolsep}{4pt}
\begin{tabular*}{\columnwidth}{@{\extracolsep{\fill}}lll@{}}
\hline
\textbf{Positions} & uniform $\bar\eta$ (U) & per-pinch (P) \\
\hline
equal spacing (E) & EU & EP \\
above user (A) & AU & AP \\
optimized (O) & OU & OP \\
\hline
\textbf{Off-grid scheme} & \textbf{starts from} & \textbf{and changes} \\
\hline
OP-seq & OU & efficiencies re-solved per pinch \\
EU/OP-hedged & EU/OP & \eqref{eq:C1} for every PR location \\
OP-free & OP & \eqref{eq:C1} removed \\
OP-inst & OP's positions & allocation each interval$^{\dagger}$ \\
H1 & OP's positions & best single pinch each interval$^{\dagger}$ \\
SAC & --- & learned policy, moving under \eqref{eq:C4}$^{\dagger}$ \\
\hline
\multicolumn{3}{@{}l}{\scriptsize $^{\dagger}$From the instantaneous desired-link gains; every other scheme uses expected gains.} \\
\end{tabular*}
\end{table}

\subsubsection{How the schemes are solved}
\label{subsubsec:etarule}
The findings are differences between the schemes. Therefore, their numerical components are evaluated under a common standard 
so that solver effort does not affect those differences. This standard has three parts.

First, for schemes with a uniform efficiency, $\bar\eta$ is the value that maximizes the solver's objective over the whole feasible set, not the largest feasible value. The two coincide when~\eqref{eq:C1} binds, which at the nominal threshold happens on all $800$ episodes for EU, $797$ for AU and $796$ for OU. When the constraint is relaxed (Section~\ref{subsec:sensitivity}), the largest feasible value approaches one, and the first pinch would then take almost all of the power. At $+20\,\mathrm{dB}$, this would cost EU, AU and OU between $0.21$ and $0.26\,\mathrm{bit/s/Hz}$.

Second, for schemes where $\veta$ allocation is convex, solutions are calculated in terms of the extraction shares from Section~\ref{subsubsec:qvars} to a first-order (Frank--Wolfe) optimality gap below $10^{-9}$ on every episode. Since the fading-averaged rate of~\eqref{eq:P1} is concave in the shares, if the schemes maximized it, the gap would bound their suboptimality. They maximize the approximation~\eqref{eq:surrogate} instead, which is not guaranteed to be concave. 
At $\kappa = 5$\,dB its Hessian is negative definite at every returned allocation, but concavity is not proven, and the outer position search is heuristic.

Third, for schemes that search over positions (OU, OP, OP-free and OP-hedged), the search procedure is identical. It begins with a sweep of $645$ clustered position patterns, which is then refined by separate Nelder--Mead runs initialized at the six best patterns, the EU and AU positions as well as twelve seeded random positions, with the same stopping rule: same iteration cap and same tolerance. As Section~\ref{subsec:region} shows, the position search takes seconds per episode, whereas simpler rules require only milliseconds.
Each of these schemes also evaluates, as extra candidates, positions found by the other schemes and by an earlier run of the joint search. In particular, OP evaluates the positions of OU and OP-hedged, so its rate is never below that of OU or EP. 
For a meaningful comparison, certain schemes inherit the positions from certain others: OP-seq inherits OU's positions, while OP-inst and H1 inherit OP's. The common search ensures that the search budget is matched across the aforementioned four schemes. Note that this does not imply equal convergence quality. 
The clustered position patterns utilized by the fixed initial sweep constitute strong starting points for OU, OP and OP-free since their best solutions naturally form clusters. They are less effective for OP-hedged, for which the best solution came from one of the random starts in $565$ of the $800$ episodes.

\subsection{Rate and the Computational Cost}
\label{subsec:region}

\begin{figure}[!t]
    \centering
    \includegraphics[width=\columnwidth]{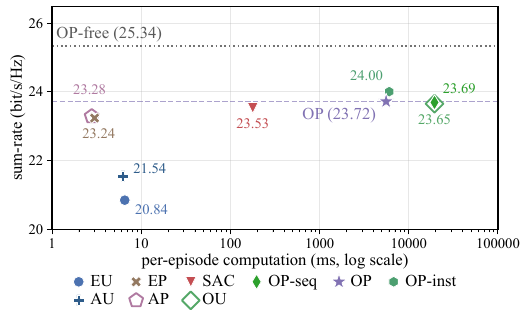}
    \caption{Sum-rate versus per-episode computation, all costs measured on one machine. SAC's cost is a full episode of $T = 200$ decisions. H1, which costs OP's search plus $14$\,ms, is in Table~\ref{tab:summary}.}
    \label{fig:ladder}
\end{figure}

In Fig.~\ref{fig:ladder}, we present a comparison of all schemes that know the PR location in terms of achievable sum-rate versus computational cost. The two hedged schemes are not included because they are about information and not computation time and are discussed in Section~\ref{subsec:adaptivity}. Table~\ref{tab:summary} summarizes performances at the nominal operating point. Three off-grid schemes attain rates above OP. This does not invalidate OP's role as a reference, since it is not stated as an upper bound. OP-free beats OP because it does not adhere to the interference constraint, whereas OP-inst and H1 use instantaneous desired-link gain information that is unavailable to OP. Therefore,  percentages above $100\%$ indicate performance relative to OP, not a violation of a bound.

\begin{table}[!t]
\renewcommand{\arraystretch}{1.25}
\caption{Sum-Rate (bit/s/Hz), Percentage of OP, and Measured Cost per Episode at the Nominal Point}
\label{tab:summary}
\centering
\footnotesize
\setlength{\tabcolsep}{3.5pt}
\begin{tabular}{@{}lrrr@{\hspace{9pt}}|@{\hspace{9pt}}lrrr@{}}
\hline
\multicolumn{4}{@{}l}{\emph{The grid}} & \multicolumn{4}{l@{}}{\emph{Off-grid schemes}} \\
\textbf{Scheme} & \textbf{Rate} & \textbf{\%} & \textbf{Cost} & \textbf{Scheme} & \textbf{Rate} & \textbf{\%} & \textbf{Cost} \\
\hline
EU & $20.84$ & $87.9$ & $6.5$\,ms & OP-seq & $23.69$ & $99.9$ & $19.5$\,s \\
EP & $23.24$ & $98.0$ & $3.0$\,ms & OP-hedged & $17.48$ & $73.7$ & $33.9$\,s \\
AU & $21.54$ & $90.8$ & $6.2$\,ms & EU-hedged & $15.54$ & $65.5$ & offline \\
AP & $23.28$ & $98.2$ & $2.8$\,ms & OP-free & $25.34$ & $106.8$ & $4.3$\,s \\
OU & $23.65$ & $99.7$ & $19.4$\,s & OP-inst & $24.00$ & $101.2$ & $6.0$\,s \\
\textbf{OP} & $\mathbf{23.72}$ & $\mathbf{100}$ & $5.6$\,s & H1 & $23.82$ & $100.4$ & $5.6$\,s \\
 & & & & SAC & $23.53$ & $99.2$ & $178$\,ms \\
\hline
\end{tabular}
\end{table}

Checking the difference between the reference (OP = $23.72$) and the unconstrained ceiling (OP-free = $25.34$) we conclude that constraint~\eqref{eq:C1} costs $1.62$, or $6.4\%$. 
The cheap schemes (EU, AU, EP, AP) attain performances between $88\%$ and $98\%$ of the reference (SAC $99\%$). EP, the scheme that makes no position adjustments, costs milliseconds and is already at $98\%$ of OP. AP reaches $23.28$ at the same cost, while the cost of the position search that provides the $2\%$ improvement is three orders of magnitude more than EP or AP and $31\times$ a full episode of SAC.

All costs were measured on the same machine, on one thread, as the mean over episode $0$ of each of the ten seeds. The unit is one whole episode, and a scheme is charged for the positions it inherits but not for the candidates it only evaluates. OP-seq includes OU's search, OP-inst and H1 include OP's, while OP is not charged for OU's positions, which it evaluates among its candidates.
For SAC, one episode is $200$ decisions of $0.89$\,ms each: $0.34$\,ms for the forward pass and $0.55$\,ms for the bisection that maps the action onto the interference budget. The costs are given only to establish the order of the schemes. They show that the position search dominates and that the efficiency allocation is cheap: OP-inst repeats the allocation $200$ times per episode and still takes only $0.4$\,s more than OP.

\subsection{What Each Degree of Freedom Is Worth}
\label{subsec:dof}

The cells of the grid that share a letter differ only in one DoF. Thus, their pairwise differences can be checked to see the contributions of positioning and radiation efficiency separately. Since it is possible to take two different paths from EU to OP and the order of these steps matters, we evaluate all four cells so that both orderings can be examined.

\begin{table}[!t]
\renewcommand{\arraystretch}{1.2}
\caption{Attribution of the Joint Gain Over Equal Spacing}
\label{tab:dof2x2}
\centering
\footnotesize
\begin{tabular}{lcc}
\hline
& uniform $\bar\eta$ & per-pinch $\veta$ \\
\hline
equal spacing      & EU \ $20.845$   & EP \ $23.235$ \\
optimized positions& OU \ $23.653$   & OP \ $23.719$ \\
\hline
\multicolumn{3}{l}{\emph{Attribution of the joint gain $+2.874$}} \\
Path A (positions first) & $+2.808$ then $+0.066$ & $42.8\times$ \\
Path B (efficiencies first) & $+2.391$ then $+0.483$ & $0.20\times$ \\
\textbf{Shapley (order-free)} & $+1.646$ vs $+1.228$ & $\mathbf{1.34\times}$ \\
Interaction & \multicolumn{2}{c}{$-2.325$} \\
\hline
\multicolumn{3}{l}{\scriptsize Ratios are computed from unrounded values.} \\
\end{tabular}
\end{table}

\begin{figure}[!t]
    \centering
    \includegraphics[width=\columnwidth]{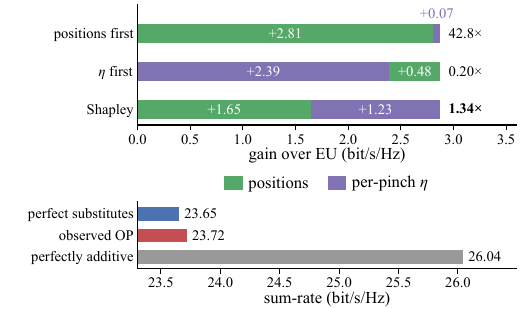}
    \caption{Attribution of the joint gain over EU (top), and the
    observed OP between the two limits (bottom). The ratio beside
    each bar is positions over efficiencies. The interaction is
    $-2.33$, and the observed OP lies $2.7\%$ of the way from
    perfect substitution to additivity.}
    \label{fig:dof}
\end{figure}

As we show in Table~\ref{tab:dof2x2} and Fig.~\ref{fig:dof}, the single-path result depends heavily on the order chosen. Repositioning, when credited first, yields $42.8\times$ the radiation efficiency, compared to just $0.20\times$ when credited second. Since only the joint total of $+2.874$ is order-invariant, we use an order-free (Shapley) attribution and calculate a ratio of $\mathbf{1.34\times}$  (with a paired-bootstrap $95\%$ interval of $1.32$--$1.36$ over the $800$ episodes). This shows that we obtain almost equal value from both DoF. 
This 
comes from the interaction term $\Delta \triangleq g_{\mathrm{joint}} - (g_p + g_\eta) = -2.325$ with a $95\%$ interval of $-2.41$ to $-2.24$, where $g_p=2.808$ and $g_\eta=2.391$ are the gains obtained when each DoF is optimized alone (EU to OU and EU to EP, respectively) and $g_{\mathrm{joint}}=2.874$ is the gain of OP over EU. This interaction constitutes 81\% of the joint gain and 45\% of the additive sum $g_p + g_\eta = 5.199$. 
The first step of Path B is the in-model analogue of relaxing the equal-split assumption of~\cite{Sun2025PASSCR, Li2026PASSCR}: letting the efficiencies vary is worth $+2.39$ at equal spacing, but only $+0.07$ once the positions are optimized.

We quantify the substitutability of the two DoF as:
\begin{equation}
    \sigma
    \triangleq
    \frac{g_{\mathrm{joint}} - \max(g_p, g_\eta)}
         {(g_p + g_\eta) - \max(g_p, g_\eta)}
    = \frac{0.066}{2.391}
    \approx 2.7\%.
    \label{eq:sigma}
\end{equation}
The numerator is what OP gains compared to the better DoF alone. The denominator is what the other DoF is worth on its own. 
Therefore, $\sigma = 0$ is perfect substitution and $\sigma = 1$ is additivity. 
The observed value of $2.7\%$ ($95\%$ interval $2.5$--$3.0\%$) is very close to perfect substitution; Section~\ref{subsec:sensitivity} shows where this comes from.

This value depends on the experiment and the signal model. Of the experiment's parameters, the threshold and the cable loss were varied. When \eqref{eq:C1} is relaxed by $10$ and $20\,\mathrm{dB}$, $\sigma$ rises to $15\%$ and $62\%$ (Section~\ref{subsec:sensitivity}). Lowering the cable loss to the $0.08$\,dB/m assumed in~\cite{Xu2026Atten, Ding2025Blockage} does the opposite: on $120$ of the $800$ episodes, drawn at random, $\sigma$ falls from $2.5\%$ to $0.6\%$, the Shapley ratio moves from $1.33\times$ to $1.42\times$, and the price of hedging at OP rises from $6.22$ to $6.34\,\mathrm{bit/s/Hz}$. The utilized position search does not have a certificate. Its effects on $\sigma$ can be bounded from below ($\sigma \ge 0$) because OP includes both single-DoF solutions (OU, EP) among its candidates, so $g_{\mathrm{joint}} \ge \max(g_p, g_\eta)$.
Even though a bound above cannot be given, a certain robustness can be claimed. For $\sigma$ to exceed the near-perfect-substitution range ($\sigma \le 10\%$), the heuristic position search would have to under-solve by at least $0.17\,\mathrm{bit/s/Hz}$ on average. This is relatively large, more than twice $0.066$, the value it found for the second DoF.

To understand the effect of the signal model on $\sigma$ we tested a coherent signal model on $120$ of the $800$ episodes, drawn at random, and compared it with the power-sum model on the same episodes. 
For these episodes we evaluated the four corners of the grid by re-solving the problem under a phase-coherent signal model. 
Here, cable permittivity has an additional role since it affects the phase with which each pinch radiates. Since a full investigation of the coherent model is out of the scope of this paper, we use a single permittivity value and a solver without a certificate, leaving the detailed study to future work.
The interference constraint~\eqref{eq:avg_interference_constraint} is on the average
interference and is averaged over the PR-path phases. Therefore, we keep the interference incoherent.
Since positions affect the received signal phases whereas efficiencies cannot, positions have a higher impact:
$g_p = 4.17$ against $g_\eta = 2.99$, and a Shapley ratio of
$1.73\times$ ($1.66$--$1.81$) against $1.34\times$ for the power-sum
model on the same episodes. The interaction stays large, $-2.72$
($-2.98$ to $-2.47$), and $\sigma$ rises from $2.5\%$ to $9.1\%$
($7.8$--$10.5\%$).
We see that the two are still strong substitutes, but now at the edge of the near-perfect-substitution range. When the constraint is relaxed by $40\,\mathrm{dB}$, $\sigma$ moves to $93\%$ under coherence as well, so Finding~\ref{find:constraint} holds. Finding~\ref{find:adapt} has not been measured coherently.

\subsubsection{What is left once both are optimized}
\label{subsubsec:opportunism}
Once both DoF are optimized, any further improvement can only be obtained by additional information, such as instantaneous fading. When it is provided, allowing adjustments every reconfiguration interval we obtain an additional $+0.286\,\mathrm{bit/s/Hz}$ (OP-inst $ = 24.005$ while OP $ = 23.719$). Although this gain is only about a tenth of $+2.874$ obtained by the joint position--efficiency optimization, it is still $4.4\times$ what radiation efficiency adds once positions are optimal. 
H1 adds only $14$\,ms per episode to OP's computation, against $0.47$\,s for OP-inst. It yields $23.825$, $0.106\,\mathrm{bit/s/Hz}$ above OP ($95\%$ interval $0.089$--$0.122$), and so captures $37\%$ of the headroom ($31$--$42\%$).
With these results, we can add to Finding~\ref{find:subst} that the value of a DoF also depends on the information available. On \emph{expected gains}, radiation efficiency on optimized positions adds $0.066$. Re-solving on \emph{instantaneous fading information}, we obtain $0.286$ more.

\subsubsection{How many pinches OP feeds}
\label{subsubsec:radiators}

The model allows up to $K$ pinches, but OP may choose not to use them all. When $\eta_k = 0$ pinch $k$ radiates no power, adds nothing to the interference 
and has no effect on the downstream pinches. Therefore, OP's search already includes solutions where one or more pinches are not fed (inactive). If we consider a pinch as inactive when it radiates less than $1\%$ of the total radiated power, then OP feeds all four pinches on $797$ of the $800$ episodes (on average $3.996$ pinches). 
To understand why OP uses all four pinches, we first check coverage, one pinch near each SU. However, checking the positions we see that OP usually packs pinches together.  On $513$ of the $800$ episodes gaps between pinches are at the minimum $0.5$\,m separation, 
where the four pinches lie within a $1.5$\,m range on the $20$\,m cable. They essentially cover one place, not four. The reason cannot be providing coverage.

Next we check fading. 
Since the rate is concave in the received power, highly fluctuating received power yields a smaller average rate. 
In the model, paths fade independently, so 
multiple radiating pinches cause smaller fluctuations compared to a single radiating pinch.
OP does not know which of the four paths has smaller fading. It also has to commit to a single allocation during the whole episode. Therefore, it spreads the radiated power to reduce the fluctuation. 
Indeed, 
when OP is modified to ignore fading (the second term of~\eqref{eq:surrogate} dropped),
it activates only $1.78$ pinches on average and never all four ($0$ out of the $800$ episodes), losing $0.129$\,bit/s/Hz. So we conclude that the reason OP feeds all pinches, even when it packs them side by side, is indeed to combat fading.

\subsection{The Price of Not Knowing the PR Location}
\label{subsec:adaptivity}

The two hedged schemes know the region where PR can be located, but not the actual PR location. Therefore, they have to conform even to the PR location that would cause the most interference. As a result, OP-hedged radiates only $0.36\%$ of the injected power on average, against $28\%$ for OP.
Since the objective is not a function of PR location (only~\eqref{eq:C1} is), the difference between a scheme and its hedged counterpart is the price of hedging against the worst-case PR location (Section~\ref{subsec:regimes}). 
EU-hedged and OP-hedged yield $15.54$ and $17.48\,\mathrm{bit/s/Hz}$, respectively (Table~\ref{tab:summary}). Consequently, the value of PR location is $5.30\,\mathrm{bit/s/Hz}$ for EU ($95\%$ interval $5.22$--$5.39$) and $6.23$ for OP ($6.19$--$6.28$).
Compared to the value of each DoF (for EU, $2.81$ and $2.39$), even compared to both DoF together ($2.87$), the value of PR knowledge is much higher, which supports Finding~\ref{find:adapt}. 

EU, OP and their hedged versions form a grid similar to the grid given in Section~\ref{subsec:dof}. Here the PR location and the optimization of both DoF are the factors. The new interaction is $+0.93$ ($0.85$--$1.02$). 
Optimizing both (OP - EU) is worth $2.87$ when the PR location is known (Table~\ref{tab:dof2x2}) and only $1.94$ when it is not (OP-hedged $-$ EU-hedged). With the DoF, interaction was negative and the two DoF were substitutes. Here, we have a positive interaction so information and optimization are complements.
Starting from EU-hedged, adding PR location (EU $-$ EU-hedged) is worth $5.30\,\mathrm{bit/s/Hz}$. From the same start, adding optimization (OP-hedged $-$ EU-hedged) is worth $1.94$. Checking the opposite order, starting from OP and removing PR knowledge (OP $-$ OP-hedged) costs $6.23$ and from the same corner removing the optimization (OP $-$ EU) costs $2.87$.
Under the given conditions PR knowledge is considerably more valuable than optimizing both DoF. 
Unlike the strong interaction between the DoF in Finding~\ref{find:subst}, ordering hardly matters here.\footnote{Before the final results we had chosen a different test of this finding: $\tau$, the gain of AU over OP-hedged divided by the gain of OP over AU, here $4.05/2.18 = 1.86$ ($95\%$ interval $1.75$--$1.98$). Its numerator combines the gain from the PR location with the optimization lost between OP-hedged and AU, so we use only the sign of $\tau$; $\tau > 1$ agrees with the comparison above.}

When computation times are examined, we see that giving up only on position search, there is a huge drop from $5.6\,\mathrm{s}$ per episode for OP to $3.0\,\mathrm{ms}$ for EP, while the drop in rate is only $0.48\,\mathrm{bit/s/Hz}$ (Fig.~\ref{fig:ladder}). If we instead give up only on PR location knowledge, OP-hedged loses $6.23$ and also \emph{loses} computation time as well. Its search takes $33.9\,\mathrm{s}$ per user layout, six times OP's. However, it can be run before the PR location is known. 
Hedging costs considerably more than any other single factor. Even against the position search, which is the second most costly factor, it loses thirteen times more rate without any gain on the computation time.

\emph{This mechanism is not specific to cognitive radio.}
The cause of the hedging cost is three features of the constraint. None of them are specific to PR. First, pinches cannot be optimized independently, because the constraint enforces a shared budget. 
Second, if certain elements are not known when choosing positions and efficiencies, the constraint must be satisfied for every possible condition.
Third, since the objective is independent of the unknown element, the gap between adaptive and hedged designs enables isolating the cost of hedging. 
Some examples of similar situations are: a health related exposure limit to protect a person at an unknown location, or a security related power cap intended for an eavesdropper at an unknown location.

\subsection{Fairness Across Users}
\label{subsec:fairness}

To see if a higher sum-rate comes at the expense of fairness, we check the $5$th-percentile per-user rate, taken over all $640\,000$ per-user samples of each scheme. We see that it rises monotonically with the sum-rate: $4.49\,\mathrm{bit/s/Hz}$ for EU, $4.58$ for AU, $5.36$ for OU and $5.42$ for OP. The $5$th-percentile per-user rate increases by 0.93 from EU to OP while the average per-user rate only increases by 0.72.
Therefore, maximizing the sum rate does not come at the expense of the weakest users. In fact, their rates gain more than the average does.

\subsection{The Interference Threshold}
\label{subsec:sensitivity}

\begin{figure}[!t]
    \centering
    \includegraphics[width=\columnwidth]{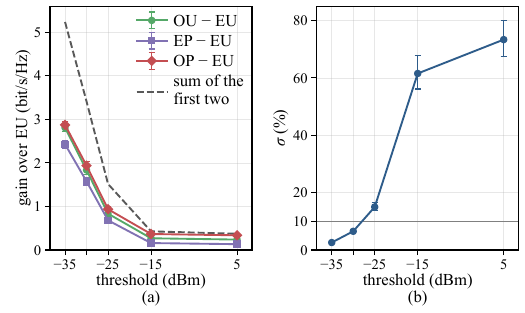}
\caption{The threshold sweep on the $800$ episodes, scored on the
approximation~\eqref{eq:surrogate}: \textbf{(a)} gains over EU,
\textbf{(b)} $\sigma$. Bars are $95\%$ intervals; the line marks $\sigma = 10\%$.}
    \label{fig:sweep}
\end{figure}

We relax the interference threshold to see if substitutability given in Section~\ref{subsec:dof} still holds. 
We repeat the grid experiment of Table~\ref{tab:dof2x2} on the same $800$ episodes at offsets of $+5$, $+10$, $+20$ and $+40\,\mathrm{dB}$ (Fig.~\ref{fig:sweep}). 
Each configuration can be scored in two ways: by the reported rate (by averaging over the $T$ fading draws of the episode), or by the approximation~\eqref{eq:surrogate} (the quantity the solver maximizes). We use the approximation here. It guarantees OP never attains a lower rate than EP and OU, since their configurations are also evaluated by OP's search. Consequently, the bound given below is a guarantee. On the other hand, the reported rate does not guarantee OP's dominance all the time. Although all schemes use the same fading draws, a certain draw may help one configuration more than another and where two schemes are close their order can reverse (on $69$ of the $800$ episodes at the nominal threshold, OU beats OP by at most $0.012\,\mathrm{bit/s/Hz}$). However, these reversals are few and small compared to an average lead of $0.066$. 
So the conclusion does not change. With the reported rate $\sigma$ is calculated as  $2.7$, $6.9$, $16$, $65$ and $78\%$ at the nominal threshold and the four offsets, respectively. With the approximation we have $2.6$, $6.5$, $15$, $62$ and $73\%$. 
Finally, the nominal interaction is given as $-2.37$ here, against the $-2.33$ reported in Table~\ref{tab:dof2x2} because the approximation scores EP
$0.04\,\mathrm{bit/s/Hz}$ higher than its reported rate.
At the respective threshold offsets, the mean interaction falls from that nominal value ($-2.37$) to $-1.47$, $-0.58$, $-0.06$ and $-0.04$. The two single-DoF gains $g_p$ and $g_\eta$ keep $8.5\%$ and $5.5\%$ of their nominal values at $+40\,\mathrm{dB}$ offset, respectively, while the interaction keeps even less than $2\%$. Interaction diminishes faster than the gains it couples.
Here, two points are worth emphasizing.
First, with such small  \emph{mean} interactions at the relaxed threshold values, ($97.4\%$ removed at $+20\,\mathrm{dB}$ ($95\%$ interval $96.8$--$98.0\%$)
and $98.5\%$ at $+40\,\mathrm{dB}$) the \emph{per-episode} interactions actually have mixed signs. The signs are positive on $429$ of the $800$ episodes at $+20\,\mathrm{dB}$ and on $455$ at $+40\,\mathrm{dB}$.
Second, recall that these percentages are obtained using the heuristic search of the OP reference. So, OP may be underestimated. With small interactions, this may make $\Delta$ artificially more negative. Therefore, we also provide a bound that is independent of the quality of the heuristic.
Since OP is evaluated at the single-DoF schemes (OU and EP), it is at least as good as them on every episode. Then $\text{OP}\,\geq\,\max(\text{OU, EP})$. So the interaction can at most be equal to the smaller one: $|\Delta| \le \min(g_p,
g_\eta)$. At every offset the smaller gain is $g_\eta$ ($0.159$ at $+20\,\mathrm{ dB}$ and $0.134$ at $+40\,\mathrm{dB}$), which does not use position search (no heuristic). 
Therefore, the bound guarantees that at least $93\%$ of the nominal interaction is removed at $+20\,\mathrm{dB}$ and $94\%$ at $+40\,\mathrm{dB}$.

Interaction being negative does not necessarily prove that the two DoF are coupled. Another reason may be that the rate is concave in the received power. We distinguish between the two as follows. We start with the idea that at high SNR, an increase of $g$ 
corresponds approximately to multiplying the received power by $2^g$. If a position optimization alone gives $g_p$, this would correspond to an increase of $2^{g_p}-1$ from the baseline power. Thus two independent mechanisms would give $1+(2^{g_p}-1)+(2^{g_\eta}-1)$. So two independent mechanisms would yield a rate increase of $\log_2(2^{g_p} + 2^{g_\eta} - 1)$.
At the nominal threshold this averages to $3.47\,\mathrm{bit/s/Hz}$, for which $\sigma=27\%$. However, the measured joint gain is $2.87\,\mathrm{bit/s/Hz}$ with $\sigma=2.6\%$, and it is lower than the benchmark in all $800$ episodes. 
The difference then, must arise from the coupling. 
At $+40\,\mathrm{dB}$, the benchmark and measured gains are $0.344$ ($\sigma=78\%$) and $0.337\,\mathrm{bit/s/Hz}$ ($\sigma=73\%$), respectively. The two are almost indistinguishable and 
at this point, the substitution largely vanishes. We conclude that substitution arises with the shared constraint.

For a designer the threshold at which the interaction changes regime is more valuable than the fully relaxed limit. Taking the interference at the equal-spaced geometry and at the nominal threshold as the baseline, $I/\Ith$ falls to $0.755$, $0.238$ and $0.003$ at $+10$, $+20$ and $+40\,\mathrm{dB}$ offsets. 
The transition occurs between $\Ith = -25$ and $-15\,\mathrm{dBm}$. 
The two DoF remain strong substitutes from the nominal to $-25\,\mathrm{dBm}$ ($\sigma = 15\%$) and are largely independent by $-15\,\mathrm{dBm}$ ($\sigma = 62\%$). 

%% file: body_section_VII_conclusion.tex
\section{Conclusion}
\label{sec:conclusion}

We characterized the two DoF of a cascaded pinching-antenna array, the positions of the pinches and their radiation efficiencies, under the interference constraint of an underlay CR network. For a single link, we derived the optimal position in closed form and showed that cable attenuation drops out of the position problem where the constraint binds (Propositions~\ref{prop:optimal_pk} and~\ref{prop:constrained_pk}). In extraction shares, the efficiency allocation at fixed positions is a convex program. One-pass decoupling can be strictly suboptimal (Proposition~\ref{prop:suboptimal}), but the sequential scheme we measured loses only $0.028\,\mathrm{bit/s/Hz}$.

\emph{Finding~\ref{find:subst}.} Neither DoF dominates. With an order-free attribution, positioning is worth $1.34\times$ radiation efficiency, while a single-path attribution gives $42.8\times$ or $0.20\times$, depending on the order. The two are near-perfect substitutes ($\sigma = 2.7\%$).

\emph{Finding~\ref{find:constraint}.} The substitutability is due to the constraint, not the cascade. Relaxing the threshold by $20$\,dB removes $97\%$ of the interaction, and the transition lies between $\Ith = -25$ and $-15$\,dBm.

\emph{Finding~\ref{find:adapt}.} Knowing the PR location matters more than optimizing both DoF. Not knowing it costs $5.30\,\mathrm{bit/s/Hz}$ at EU and $6.23$ at OP, while optimizing both DoF is worth $2.87$ when the location is known. Hedging also saves no computation. On the other hand, giving up the position search costs only $0.48\,\mathrm{bit/s/Hz}$ and reduces the computation by three orders of magnitude.

Under a coherent signal model, the two DoF remain strong substitutes, and the substitutability vanishes again when the constraint is relaxed; Finding~\ref{find:adapt} was not re-checked there. Findings~\ref{find:constraint} and~\ref{find:adapt} depend only on the structure of the constraint: 
a shared budget whose active constraint is unknown when the configuration is chosen and does not affect the objective. 
This structure carries over beyond cognitive radio, although the magnitudes need not.

Two questions remain open. On expected gains, OP feeds all four pinches to average out the fading (Section~\ref{subsubsec:radiators}); whether a number of pinches different from the number of users pays is open. And since the position search is what separates the inexpensive schemes from OP, an inexpensive position refiner is the main open problem. Natural extensions are non-uniform slots, multiple primary receivers, an outage-based constraint, and multi-feed precoding.